\documentclass[%
 reprint,
superscriptaddress,
nobibnotes,
amsmath,amssymb,
aps,
longbibliography
]{revtex4-2}

\usepackage{graphicx}
\usepackage{dcolumn}
\usepackage{bm}
\usepackage{hyperref}
\hypersetup{
     colorlinks   = true,
     linkcolor    = blue,
     citecolor    = blue,
}

\begin{document}

\preprint{APS/123-QED}

\title{Intermediate polaronic charge transport in organic crystals\\from a many-body first-principles approach}

\author{Benjamin K. Chang}
\author{Jin-Jian Zhou}%
\author{Nien-En Lee}
\author{Marco Bernardi}
\affiliation{Department of Applied Physics and Materials Science, California Institute of Technology, Pasadena, California 91125, USA}

\begin{abstract}
\noindent
Predicting the electrical properties of organic molecular crystals (OMCs) is challenging due to their complex crystal structures and electron-phonon (e-ph) interactions. Charge transport in OMCs is conventionally categorized into two limiting regimes $-$ band transport, characterized by weak e-ph interactions, and charge hopping due to localized polarons formed by strong e-ph interactions. 
However, between these two limiting cases there is a less well understood intermediate regime where polarons are present but transport does not occur via hopping. Here we show a many-body first-principles approach that can accurately predict the carrier mobility in OMCs in the intermediate regime and shed light on its microscopic origin. 
Our approach combines a finite-temperature cumulant method to describe strong e-ph interactions with Green-Kubo transport calculations. We apply this parameter-free framework to naphthalene crystal, demonstrating electron mobility predictions within a factor of 1.5$-$2 of experiment between 100$-$300~K.
Our analysis reveals that electrons couple strongly with both inter- and intramolecular phonons in the intermediate regime, as evidenced by the formation of a broad polaron satellite peak in the electron spectral function and the failure of the Boltzmann equation. 
Our study advances quantitative modeling of charge transport in complex organic crystals. 
\vspace{-10pt}
\end{abstract}
\maketitle
%
%
\section*{\label{sec1:intro}Introduction}
\vspace{-15pt}
Organic molecular crystals (OMCs) possess complex crystal structures with an intricate interplay of electronic and structural degrees of freedom. Their electronic properties range from metallic to semiconducting or insulating, and they can host ferroelectricity~\cite{horiuchi_organic_2008}, magnetism~\cite{miller_organic-_2014} and superconductivity~\cite{jerome_organic_2012}.
OMCs are also versatile materials with broad applications, 
for example in electronics~\cite{Facchetti}, light-emitting diodes~\cite{chen_liquid_2018,pimputkar_prospects_2009}, spintronics~\cite{dediu_spin_2009,joshi_spintronics_2016}, batteries~\cite{armand_building_2008,mauger_recent_2019} and solar cells~\cite{inganas_organic_2018,mazzio_future_2015,abdulrazzaq_organic_2013}.
The charge carrier mobility is a key figure of merit for organic materials in these devices~\cite{troisi_charge_2011,fratini_map_2017,fratini_charge_2020,nematiaram_largest_2020}. Yet, understanding charge transport and accurately predicting the mobility in OMCs remain open challenges. 
Due to the presence of electron-phonon (e-ph) interactions ranging from weak to strong~\cite{lee_charge_2018}, the mobility varies dramatically among different OMCs, both in magnitude and temperature dependence~\cite{fratini_charge_2020}. Even in the same organic crystal, electron and hole carriers can exhibit different transport regimes, and the mobility can vary by orders of magnitude for different crystallographic directions.
\\
\indent 
Charge transport in OMCs is often classified into two limiting cases $-$ the band transport and polaron hopping regimes, each entailing specific transport mechanisms~\cite{oberhofer_charge_2017}. In band transport, charge carriers are delocalized, the e-ph coupling is weak, and the mobility is correspondingly high ($>\,$10~cm$^2$/Vs) and characterized by a power-law decrease with temperature. Band transport in OMCs is usually governed by scattering of carriers with low-energy acoustic and intermolecular phonons, with the corresponding e-ph interactions often modeled by the Peierls Hamiltonian~\cite{su_solitons_1979}. These weak e-ph interactions can be described with lowest-order perturbation theory, and the OMC mobility can be accurately predicted using the Boltzmann transport equation (BTE)~\cite{lee_charge_2018,brown-altvater_band_2020}, including all phonon modes on the same footing. 

In the polaron hopping regime, the charge carriers interact strongly with phonons, forming self-localized (small) polarons, which are often modeled with the Holstein Hamiltonian to describe intramolecular e-ph interactions~\cite{holstein_studies_1959}. 
The resulting charge transport is dominated by thermally activated polaron hopping and is often described with Marcus theory~\cite{marcus_chemical_1964,fratini_transient_2016}. 
The mobility in the polaron hopping regime is relatively small, usually below 0.1 cm$^2$/Vs, and is more challenging to predict from first principles.
\\
\indent
Between these two limiting scenarios, OMCs also exhibit an \textit{intermediate} transport regime, for which neither the band transport nor the polaron hopping pictures are fully adequate~\cite{oberhofer_charge_2017,fratini_transient_2016}. 
In the intermediate regime, the mobility exhibits a bandlike power-law temperature dependence~\cite{zhou_predicting_2019,mishchenko_mobility_2015,fratini_transient_2016}, yet polarons can be present and low mobility values ($<\,$1 cm$^2$/Vs) are common~\cite{oberhofer_charge_2017}. A signature of intermediate transport is the violation of the Mott-Ioffe-Regel limit~\cite{hussey__universality_2004}, whereby the carrier mean-free-paths become smaller than the intermolecular distance~\cite{fratini_transient_2016}, making the BTE description inadequate.
\\
\indent
Various approaches have been employed to study intermediate transport in OMCs; they typically employ a Holstein Hamiltonian or a (Peierls-type) dynamical disorder Hamiltonian, or a combination of both, and obtain the mobility via linear-response theory~\cite{ortmann_theory_2009,ortmann_charge_2010,fetherolf_unification_2020,wang_roles_2007}, diffusion simulation~\cite{troisi_charge-transport_2006,wang_multiscale_2010}, surface hopping method~\cite{wang_flexible_2013}, or transient localization calculation~\cite{ciuchi_transient_2011,nematiaram_practical_2019}. These methods are highly valuable for studies of OMCs, although they usually rely on simplifying assumptions such as including only specific phonon modes and e-ph interactions, or fitting model parameters to experiments. To date, first-principles approaches to predict charge transport in the intermediate regime with quantitative accuracy are scarce, especially within rigorous treatments based on many-body perturbation theory.
\\ 
\indent
%
%
In this work, we develop rigorous calculations of the mobility in the intermediate charge transport regime in OMCs. Focusing on naphthalene crystal as a case study, we employ a finite-temperature cumulant approach~\cite{zhou_predicting_2019} to capture the strong e-ph interactions and polaron effects characteristic of the intermediate regime, and employ Green-Kubo theory to compute the electron mobility. All phonon modes are included and treated on equal footing. 
This cumulant plus Kubo (CK) approach is shown to predict the electron mobility in the intermediate regime with a high accuracy, within a factor of two of experiment between 100$-$300~K for crystallographic directions parallel to the naphthalene molecular planes. We additionally show the failure of the BTE to describe the mobility in the intermediate regime.
\\
\indent
Our analysis of the electronic spectral functions reveals the presence of a broad satellite next to the quasiparticle (QP) peak, explaining the breakdown of the BTE and the band transport picture. Although the polaron satellite peak includes contributions from both inter- and intramolecular phonons, we find that the mobility is mainly limited by low-energy intermolecular phonons, 
similar to the band transport regime.
For charge transport normal to the molecular planes, we find that both the BTE and CK approaches cannot correctly predict the mobility, which experimentally is nearly temperature independent and governed by small-polaron hopping~\cite{madelung_naphthalene_nodate,schein_band-hopping_1979,schein_observation_1978}. 
This finding restricts the applicability of the CK method to intermediate e-ph coupling strengths. 

Taken together, our work provides an accurate first-principles method to study polaron transport in OMCs, and unravels the interplay of low- and high-energy phonon modes in the intermediate regime. 
Our results provide a blueprint for studying charge transport in a wide range of organic crystals. 

\section*{Results}
\vspace{-5pt}
\subsection*{\label{sec:computationalmethod} \normalsize Computational approach}
\vspace{-10pt}
We compute the ground state electronic structure of naphthalene crystal using plane-wave density functional theory (DFT) calculations with the \textsc{Quantum Espresso} code~\cite{giannozzi_quantum_2009,giannozzi_advanced_2017}. 
We employ the generalized gradient approximation~\cite{perdew_generalized_1996} and norm-conserving pseudopotentials~\cite{troullier_efficient_1991} from Pseudo Dojo~\cite{van_setten_pseudodojo_2018}. The DFT band structure is refined using GW calculations (with the {\sc Yambo} code~\cite{hedin_new_1965,marini_yambo_2009}) to better capture dynamical screening effects. 
Maximally localized Wannier functions~\cite{marzari_maximally_2012} are generated with the \textsc{Wannier90} code~\cite{mostofi_wannier90_2008} following a procedure similar to Ref.~\cite{lee_charge_2018}. We compute the e-ph interactions and charge transport separately at four temperatures (100, 160, 220, and 300~K), using different experimental lattice parameters at each temperature~\cite{brock_temperature_nodate} and relaxing the atomic positions with DFT. We obtain the lattice dynamics and e-ph perturbation potentials from density functional perturbation theory (DFPT)~\cite{baroni_phonons_2001}, and compute the e-ph interactions with the \textsc{Perturbo} code~\cite{zhou_perturbo_2020}.
%
Additional numerical details are provided in the Methods section.
\\
\indent
Using the computed e-ph interactions, we study charge transport in the BTE and CK frameworks with the~\textsc{Perturbo} code~\cite{zhou_perturbo_2020}. In the BTE, the mobility tensor $\mu_{\alpha\beta}$ is computed in the relaxation time approximation (RTA):
\begin{eqnarray}
\label{eq:mu_BTE}
\mu_{\alpha\beta}(T)=\frac{2e}{n_cV_{\text{uc}}}\int &&dE\left(-\frac{\partial f(E,T)}{\partial E}\right)\nonumber\\ &&\times\sum_{n\mathbf{k}}\tau_{n\mathbf{k}}(T)v_{n\mathbf{k}}^{\alpha}v_{n\mathbf{k}}^{\beta}\delta(E-\varepsilon_{n\mathbf{k}}),
\end{eqnarray}
where $\alpha$ and $\beta$ are Cartesian directions parallel to the crystal principal axes, $T$ is the temperature,  $e$ the electronic charge, $n_c$ the carrier concentration, $V_\mathrm{uc}$ the unit cell volume, $f$ the electronic Fermi-Dirac distribution and $E$ is the electron energy. Here and below, $n$ is the band index and $\mathbf{k}$ the crystal momentum of the electronic states. 
The BTE mobility depends on the electron band energies $\varepsilon_{n\mathbf{k}}$, the corresponding band velocities $v_{n\mathbf{k}}$, and the state-dependent e-ph relaxation times $\tau_{n\mathbf{k}}$ obtained within lowest-order perturbation theory~\cite{bernardi_first-principles_2016,zhou_perturbo_2020}.
As a sanity check, we compute the mobility at 220~K by solving the full linearized BTE with an iterative approach (ITA)~\cite{zhou_perturbo_2020,li_electrical_2015}, and find that in naphthalene it gives results identical to the RTA, justifying our use of the RTA. 
\\
\indent
To properly treat strong e-ph interactions and include polaron effects in the mobility, we employ a finite-temperature cumulant approach in which the retarded electron Green's function $G^{\mathrm{R}}_{n\mathbf{k}}$ is written using the exponential ansatz~\cite{kas_cumulant_2014,zhou_predicting_2019,mahan,aryasetiawan_gw_1998,story_cumulant_2014,nery_quasiparticles_2018}
\begin{eqnarray}
\label{eq:green}
G^{\mathrm{R}}_{n\mathbf{k}}(t,T)=G^{\mathrm{R,0}}_{n\mathbf{k}}(t)e^{C_{n\mathbf{k}}(t,T)},
\end{eqnarray}
where $G^{\mathrm{R,0}}_{n\mathbf{k}}$ is the non-interacting Green's function and $C_{n\mathbf{k}}$ is the cumulant function, obtained here at finite temperatures from the lowest-order e-ph self-energy (see Methods). The electron spectral function is obtained from the Green's function at each electron energy $E$ using
\begin{eqnarray}
\label{eq:spectral}
A_{n\mathbf{k}}(E,T)=-\mathrm{Im} G^{\mathrm{R}}_{n\mathbf{k}}(E,T)/\pi.
\end{eqnarray}
In the CK method, the mobility tensor is computed directly from the spectral function using the linear-response Green-Kubo formula~\cite{zhou_predicting_2019,economou_greens_2006_ch8, mahan}:
\begin{eqnarray}
\label{eq:mu_CK}
\mu_{\alpha\beta}(T)=\frac{1}{n_ce}\int\! dE\, \Phi_{\alpha\beta}(E,T),
\end{eqnarray}
where the integrand is the transport distribution function (TDF). Under the approximation of neglecting vertex corrections, the TDF reads~\cite{economou_greens_2006_ch8} 
\begin{eqnarray}\label{eq:TDF}
\Phi_{\alpha\beta}(E,T)=\frac{\pi\hbar e^2}{V_{\text{uc}}}\sum_{n\mathbf{k}}v_{n\mathbf{k}}^{\alpha}v_{n\mathbf{k}}^{\beta}|A_{n\mathbf{k}}(E,T)|^2\left(-\frac{\partial f(E,T)}{\partial E}\right)\!,\nonumber\\
\end{eqnarray}
where $v_{n\mathbf{k}}$ are the unperturbed electron band velocities, the same as those used in Eq.~(\ref{eq:mu_BTE})~\cite{economou_greens_2006_ch8}.
The CK mobility defined in Eq.~(\ref{eq:mu_CK}) is obtained from the cumulant spectral function, therefore it takes into account the strong e-ph coupling and polaron effects. The CK calculations have been shown to provide results in close agreement with the BTE-RTA in the limit of weak e-ph interactions (see Ref.~\cite{zhou_predicting_2019} for a calculation on GaAs).\vspace{5 pt}\\

\subsection*{\label{sec:mobility} \normalsize  Electron mobility}
\vspace{-10pt}
\begin{figure}[tb!]
\includegraphics[width=0.49\textwidth]{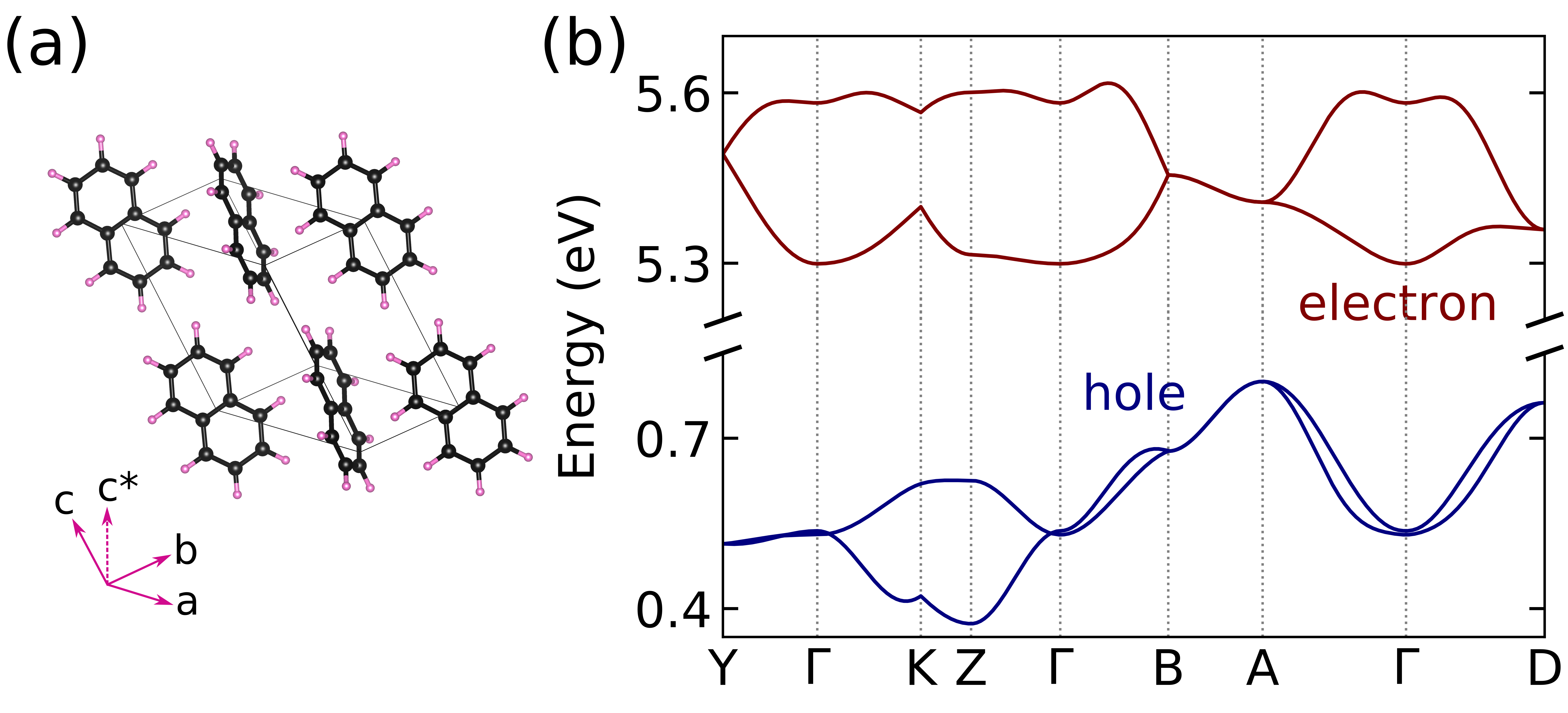}
\caption{(a) Monoclinic crystal structure of naphthalene, with molecular $a$-$b$ planes stacked in the plane-normal $c^*$ direction. (b) Band structure of naphthalene, showing the two lowest-energy electron (LUMO and LUMO+1) and hole (HOMO and HOMO-1) bands.}
\label{fig:structure} 
\end{figure} 
The crystal structure of naphthalene consists of molecular planes in the $a$ and $b$ crystallographic directions, stacked along the plane-normal $c^*$ direction [Fig.~\ref{fig:structure}(a)]. We first discuss charge transport in the molecular planes. For hole carriers, we have previously shown that such in-plane transport is bandlike and well described by the BTE~\cite{lee_charge_2018}. 
In this work, we focus on the mobility of the \textit{electron} carriers, which due to their flatter electronic bands with greater effective masses compared to holes [Fig.~\ref{fig:structure}(b)] are expected to exhibit lower mobilities and a range of transport regimes. Only the electronic bands formed by the lowest unoccupied molecular orbital (LUMO) and the next-higher-energy orbital (LUMO+1) contribute to electron transport in the 100$-$300~K temperature range studied here, so we consider only these two bands in our mobility calculations. 
\begin{figure}[tb!]
\includegraphics[width=0.49\textwidth]{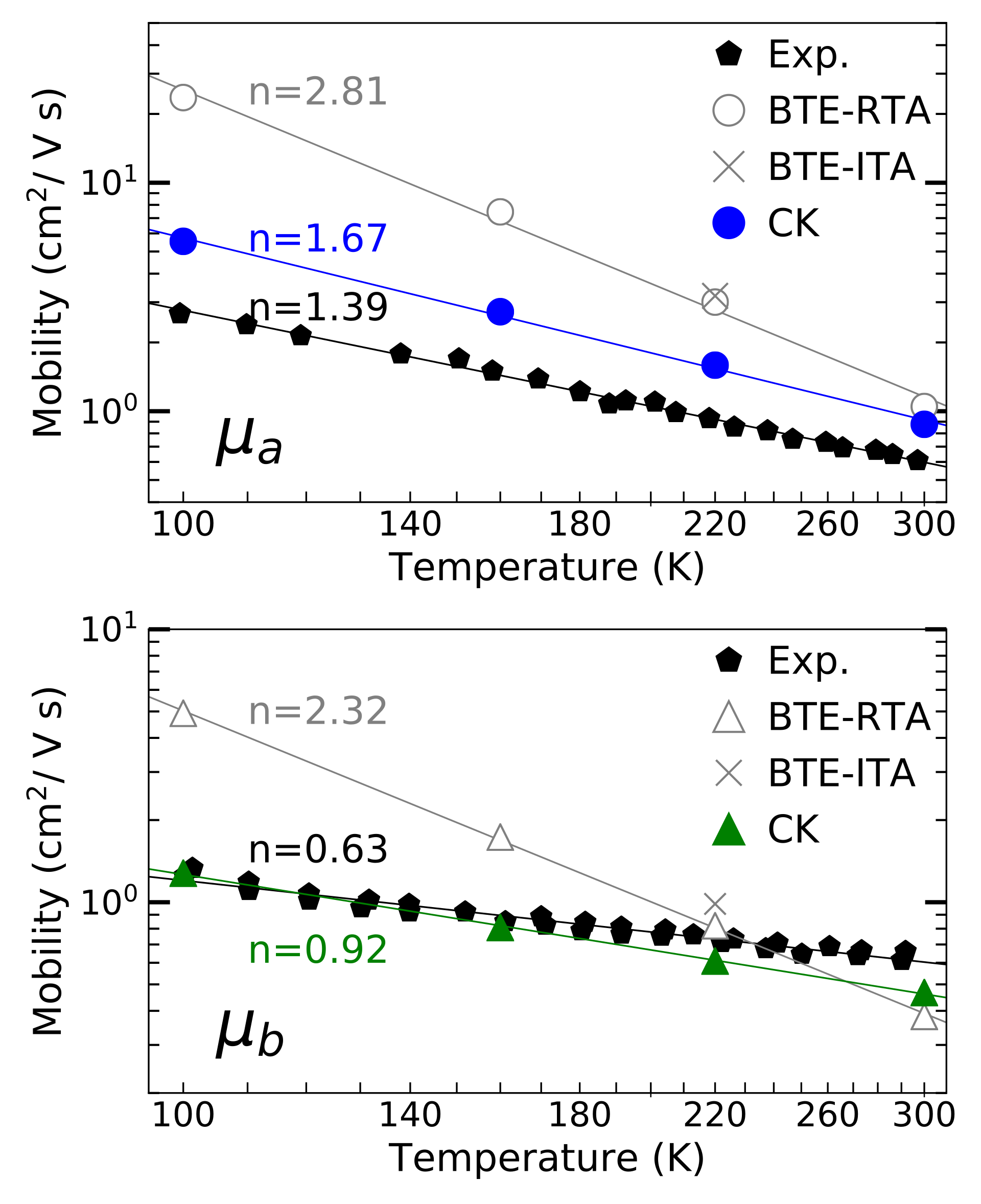}
\caption{\label{fig:mobility} Electron mobility in the in-plane directions $a$ (top panel) and $b$ (bottom panel) in naphthalene. Results obtained from BTE and CK calculations are compared with experimental data from Ref.~\cite{madelung_naphthalene_nodate}.}
\end{figure}
\\
\indent
Figure~\ref{fig:mobility} shows the in-plane electron mobilities computed with the BTE and CK methods, and compares them with experimental data~\cite{madelung_naphthalene_nodate}. 
We fit each mobility curve with a $T^{-n}$ power-law temperature trend and give the exponent $n$ next to each curve. The results show that the BTE predicts a much stronger temperature dependence of the mobility than in experiment, with errors in the computed exponents for transport along the $a$ and $b$ crystallographic directions (mobilities $\mu_a$ and $\mu_b$ in Fig.~\ref{fig:mobility}, respectively) of over 100\% for $\mu_a$ and 270\% for $\mu_b$ relative to the exponent $n$ obtained by fitting the experimental results. 
Due to this error, the BTE greatly overestimates the mobility at low temperatures $-$ for example, $\mu_a$ at 100~K from the BTE is an order of magnitude greater than the experimental value.
\\
\indent
These results are a strong evidence of the failure of the Boltzmann equation to describe electron transport in naphthalene; the physical origin of this failure is  examined below. Note that the BTE failure is not a consequence of our use of the RTA, as the full solution of the BTE~\cite{zhou_perturbo_2020} gives results nearly identical to the RTA (see the ITA points at 220~K in Fig.~\ref{fig:mobility}). The fact that the mobility has a power-law temperature dependence but is not correctly predicted by the BTE is a hallmark of the intermediate transport regime~\cite{mishchenko_mobility_2015,fratini_transient_2016}.
\begin{figure*}[htb!]
\includegraphics[width=\textwidth]{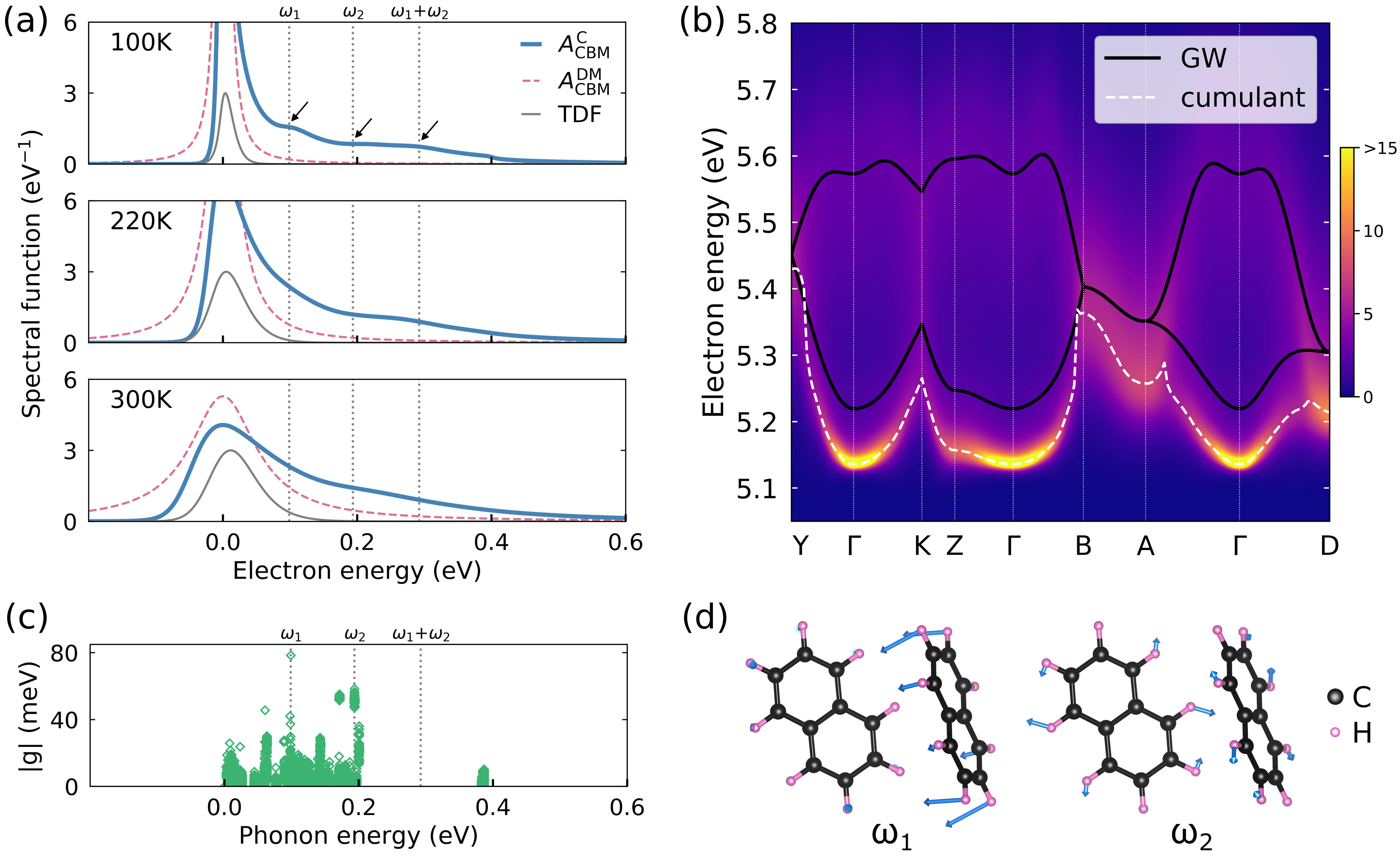} 
\caption{\label{fig:cbmspectral} (a) Spectral functions computed at three temperatures for the CBM electronic state. Results from the cumulant approach ($A^{\mathrm{C}}_{\mathrm{CBM}}$) are compared to the Dyson-Migdal spectral function ($A^{\mathrm{DM}}_{\mathrm{CBM}}$). The QP peak is chosen as the zero of the energy axis for each spectral function. The transport distribution function (TDF) in arbitrary units is also shown at each temperature. (b) Electron spectral function for the LUMO and LUMO+1 bands along a high-symmetry path, computed at 100~K using the cumulant method. The solid line is the GW band structure and the dashed line shows the renormalized cumulant band structure obtained by connecting the QP peaks of the spectral functions. 
(c) Gauge-invariant e-ph coupling strength as a function of phonon energy. The energies $\omega_1$ and $\omega_2$ of the two phonon modes with strongest e-ph coupling are shown with vertical dashed lines. (d) Atomic displacements for the two intramolecular modes with the strongest e-ph coupling.}
\end{figure*}
\\
\indent
The CK calculations give significantly improved results (Fig.~\ref{fig:mobility}). The CK mobilities are within a factor of 2 of experiment for $\mu_a$ and 1.3 for $\mu_b$ in the entire 100$-$300~K temperature range. The error in the $T^{-n}$ exponent is reduced to 20\% for $\mu_a$ and 45\% for $\mu_b$ relative to experiment, a five-fold improvement in accuracy over the BTE results. 
Achieving this level of accuracy for quantitative predictions of the mobility in OMCs has recently become possible in the band transport regime~\cite{lee_charge_2018} but has so far remained challenging in the intermediate regime. 
As we discuss below, by combining the cumulant and Green-Kubo frameworks, our CK approach can capture key polaron effects in the intermediate regime such as higher-order e-ph coupling and spectral weight transfer, resulting in improved mobility predictions. 
%
\subsection*{\label{sec:spectral} \normalsize Electron spectral function}
\vspace{-15pt}
The electron spectral function is central to understanding polaron effects~\cite{zhou_predicting_2019} and intermediate charge transport. 
The spectral function can be viewed as the density of states of a single electronic state, and it integrates to one over energy due to a well-known sum rule~\cite{mahan}. 
In Fig.~\ref{fig:cbmspectral}(a), we show the spectral function at three temperatures, using results obtained with our cumulant method for the electronic state at the conduction band minimum (CBM) [$\Gamma$ point in Fig.~\ref{fig:structure}(b)]. 
At 100~K, next to the main QP peak we find a broad spectral feature associated with the combined excitation of an electron QP plus one or two phonons. This broad satellite combines contributions from multiple satellite peaks, as shown by the arrows in Fig.~\ref{fig:cbmspectral}(a), and is a signature of polaron formation~\cite{zhou_predicting_2019}. At higher temperatures, the QP and  satellite peaks broaden and ultimately merge into a continuum at 300~K. The coexistence of a well-formed QP peak and broad satellites shows that large-polaron effects, characteristic of e-ph interactions with intermediate strength, are a key characteristic of the intermediate transport regime.
\\
\indent
The cumulant spectral functions for multiple electronic states in the LUMO and LUMO+1 bands can be combined to obtain a polaron band structure renormalized by the e-ph interactions. Figure~\ref{fig:cbmspectral}(b) compares the band structures at 100~K computed with the GW method and with our cumulant calculations that use the GW band structure as input. The cumulant band structure, obtained by connecting the QP peaks of the cumulant spectral functions at neighboring $\mathbf{k}$-points, captures polaron effects such as QP mass and weight renormalization. 
At 100~K, where the QP peaks are well-defined, we find that the renormalized effective masses in the cumulant band structure are greater by 15$-$35\% than in GW for the in-plane directions.  This shows that the cumulant approach can capture the band-narrowing due to strong e-ph interactions and polaron effects in OMCs~\cite{hannewald_theory_2004}.
\\
\indent 
The physical origin of the polaron satellite in Fig.~\ref{fig:cbmspectral}(a) is of key importance. In the prototypical case of a polar inorganic material with strong e-ph coupling with longitudinal optical (LO) phonons, the satellite peaks are located at the LO-mode energy $\omega_{\mathrm{LO}}$ (and its multiples) relative to the QP peak~\cite{zhou_predicting_2019,zhou_ab_initio_2021}. 
Here, due to the presence of a large number of phonon modes in OMCs (108 in napthalene), the satellites merge into a broad spectral feature resembling a long tail of the QP peak, with contributions from various phonon modes. To explain the origin of this broad satellite, in Fig.~\ref{fig:cbmspectral}(c) we analyze the e-ph coupling strength for an electronic state near the CBM, as quantified by the absolute value of the gauge-invariant e-ph coupling, $|g|$  (see Methods). 
\\
\indent
In naphthalene, the 12 lowest-energy phonon modes are intermolecular, and the remaining 96 modes are intramolecular vibrations~\cite{lee_charge_2018}. In Fig.~\ref{fig:cbmspectral}(a), the mode with the strongest e-ph coupling, an intramolecular phonon with energy $\omega
_1\approx0.1$~eV, generates a satellite peak in the spectral function at energy $\omega_1$ relative to the QP peak. 
The intramolecular phonon with the second strongest e-ph coupling, with energy $\omega_2\approx0.2$~eV, gives a second contribution to the broad satellite, followed by a plateau at higher energy. 
Finally, the inflection point in the spectral function at energy $\omega_1+\omega_2$ is due to higher-order e-ph coupling from the two modes with energies $\omega_1$ and $\omega_2$. 
%
The atomic displacements associated with these two intramolecular modes are shown in Fig.~\ref{fig:cbmspectral}(d). Both modes involve vibrations of the hydrogen atoms, in one case in the carbon ring plane and in the other case normal to the carbon rings. Our analysis demonstrates that these intramolecular phonons are responsible for the formation of polarons in naphthalene. This strong coupling with intramolecular phonons and the associated satellite peak in the spectral function are consistent with recent results from the Holstein-Peierls model~\cite{fetherolf_unification_2020}. 
%
\\
\indent
Interestingly, lowest-order theory is wholly inadequate to describe this polaronic regime with intermediate e-ph coupling strength. To show this point, we compute the Dyson-Migdal (DM) spectral function [see Eq.~(\ref{eq:dyson})], which is obtained from the lowest-order e-ph self-energy and therefore does not include polaron effects. 
From the comparison of the cumulant and DM spectral functions in \mbox{Fig.~\ref{fig:cbmspectral}(a)} it is clear that the DM spectral functions have a Lorentzian shape and lack any satellite structure. As a result, the subtle interplay between inter- and intramolecular phonons in the QP and satellite peaks cannot be captured in lowest-order theory. As we discuss below, this is the origin of the failure of the BTE to describe transport in the intermediate regime.
%
%
\begin{figure}[t!]
\includegraphics[width=0.36\textwidth]{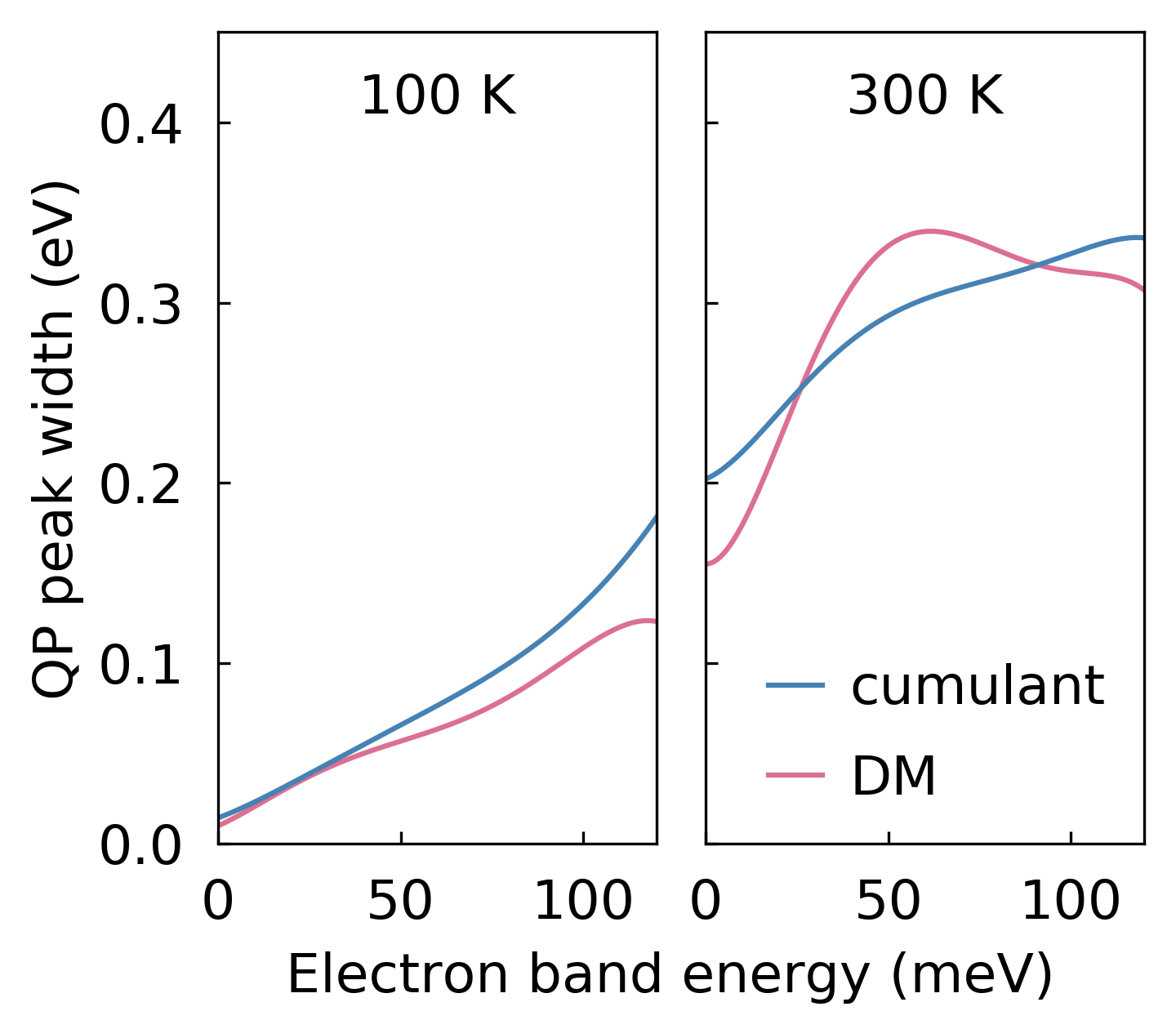}
\caption{Full width at half maximum of the QP peak, shown at two temperatures for both the cumulant and DM spectral functions. The energy zero is set to the conduction band minimum.}
\label{fig:width} 
\end{figure}
\subsection*{\mbox{\normalsize Failure of the Boltzmann equation}}
\vspace{-10pt}
It is important to understand the microscopic origin of the failure of the BTE, and the success of the CK method, to describe transport in the intermediate regime. 
In the Green-Kubo framework, the mobility is given by an integral over electron energies [see Eq.~(\ref{eq:mu_CK})], which in principle combines contributions from all features of the spectral function. To quantify the contributions of the QP and satellite peaks to charge transport, we analyze the mobility integrand, the TDF in Eq.~(\ref{eq:TDF}), and plot it together with the spectral functions in Fig.~\ref{fig:cbmspectral}(a). We find that the TDF decays rapidly outside the QP peak, within an energy $\omega_1$ of the QP peak at low temperature and $\omega_2$ at 300~K. Therefore any spectral function feature with energy greater than $\omega_2$ does not overlap with the TDF and cannot contribute to charge transport between 100$-$300~K. 
In this temperature range, although the mobility is mainly governed by the QP peak, polaron effects still contribute in important ways. 
\begin{figure*}[htb!]
\includegraphics[width=\textwidth]{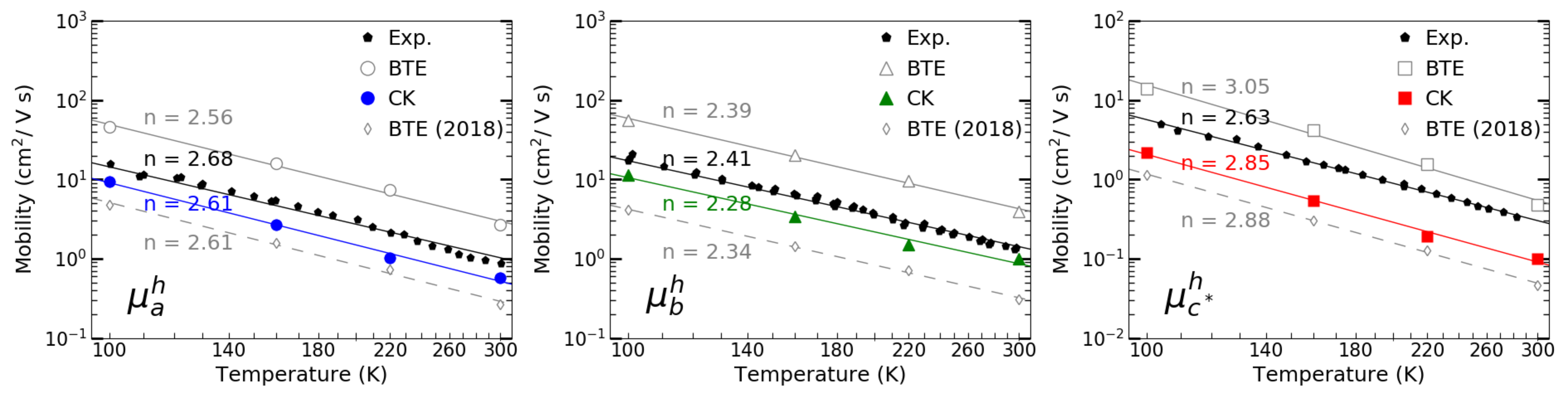}
\caption{\label{fig:holemobilty} Hole carrier mobilities ($\mu^h$) in naphthalene in the three crystallographic directions $a$, $b$, and $c^*$, computed both with the BTE in the relaxation time approximation (gray solid line) and with the CK approach (blue, green and red). The computational settings are the same as in the electron mobility calculations. The BTE results presented in Ref.~\cite{lee_charge_2018}, which were obtained using a previous implementation of e-ph interactions with less accurate Wigner-Seitz cell summations, are shown as dashed lines for comparison. The exponent $n$ from a $T^{-n}$ power-law fit of the temperature dependence is given next to each curve. The experimental data (black) are from Ref.~\cite{madelung_naphthalene_nodate}.}
\end{figure*} 
%

First, due to higher-order e-ph coupling with both inter- and intramolecular phonons, the linewidth of the QP peak in the cumulant spectral function is different than in the DM spectral function from lowest-order theory (see Fig.~\ref{fig:width}), whose linewidth is the scattering rate entering the BTE mobility calculation. 
%
This QP linewidth discrepancy is temperature and energy dependent (Fig.~\ref{fig:width}), which explains why the BTE cannot correctly predict the value and temperature dependence of the mobility in the intermediate regime, corroborating our results in Fig.~\ref{fig:mobility}.
%
%
Second, the broad satellite in the cumulant spectral function limits the carrier mobility indirectly, by transferring spectral weight away from the QP peak (recall that the spectral function integrates to one over energy). The satellite peak at $\omega_1$ contributes directly to transport only above $\sim$200~K, where the QP peak broadens, merging with the satellite and overlapping with the TDF. 
%
\\
\indent
The picture that emerges is that electron transport in the naphthalane molecular planes is mainly governed by the scattering of QPs with renormalized weight, which couple directly with intermolecular and indirectly (via weight transfer) with higher-energy intramolecular phonons. The latter can also contribute directly to charge transport as the temperatures increases above 200~K. The ability of our CK approach to address these subtle e-ph interactions enables accurate predictions of the mobility and its temperature dependence in the intermediate regime, where the BTE with lowest-order e-ph coupling fails to capture these essential polaron effects.

%
\subsection*{\mbox{\normalsize Comparison with hole mobilities}}
\vspace{-10pt}
Finally, we present mobility results for hole carriers in naphthalene to contrast their behavior with electron carriers. We compute the hole mobility in naphthalene between 100$-$300~K using the CK approach, and compare the results to BTE calculations and experiments (see Fig.~\ref{fig:holemobilty}). The BTE calculations are a refinement of those we presented in Ref.~\cite{lee_charge_2018}, obtained here using a more accurate Wigner-Seitz cell summation procedure, as implemented in \textsc{Perturbo} and described in detail in Ref.~\cite{zhou_perturbo_2020}. The revised BTE mobilities follow an identical temperature trend as in our previous results in Ref.~\cite{lee_charge_2018}, but their value is now greater than experiment, a physically meaningful trend for phonon-limited mobilities.
\\
\indent
For hole carriers, both the BTE and CK methods give accurate predictions of the hole mobility, within a factor of 2$-$3 of experiment at all temperatures. The temperature dependence is nearly identical for the CK and the BTE mobilities, as shown by fitting the mobility curves with a $T^{-n}$ power-law and giving the exponent $n$ next to each curve in Fig.~\ref{fig:holemobilty}. These findings demonstrate that the band transport picture of the BTE, which is inadequate for electron carriers, is sufficient to describe transport for hole carriers due to their more dispersive bands (see Fig.~\ref{fig:structure}) and overall weaker e-ph coupling (see Fig. S1 in the Supplementary Information).
\section*{\label{sec:out} Discussion}
\begin{figure}[tb!]
\includegraphics[width=0.5\textwidth]{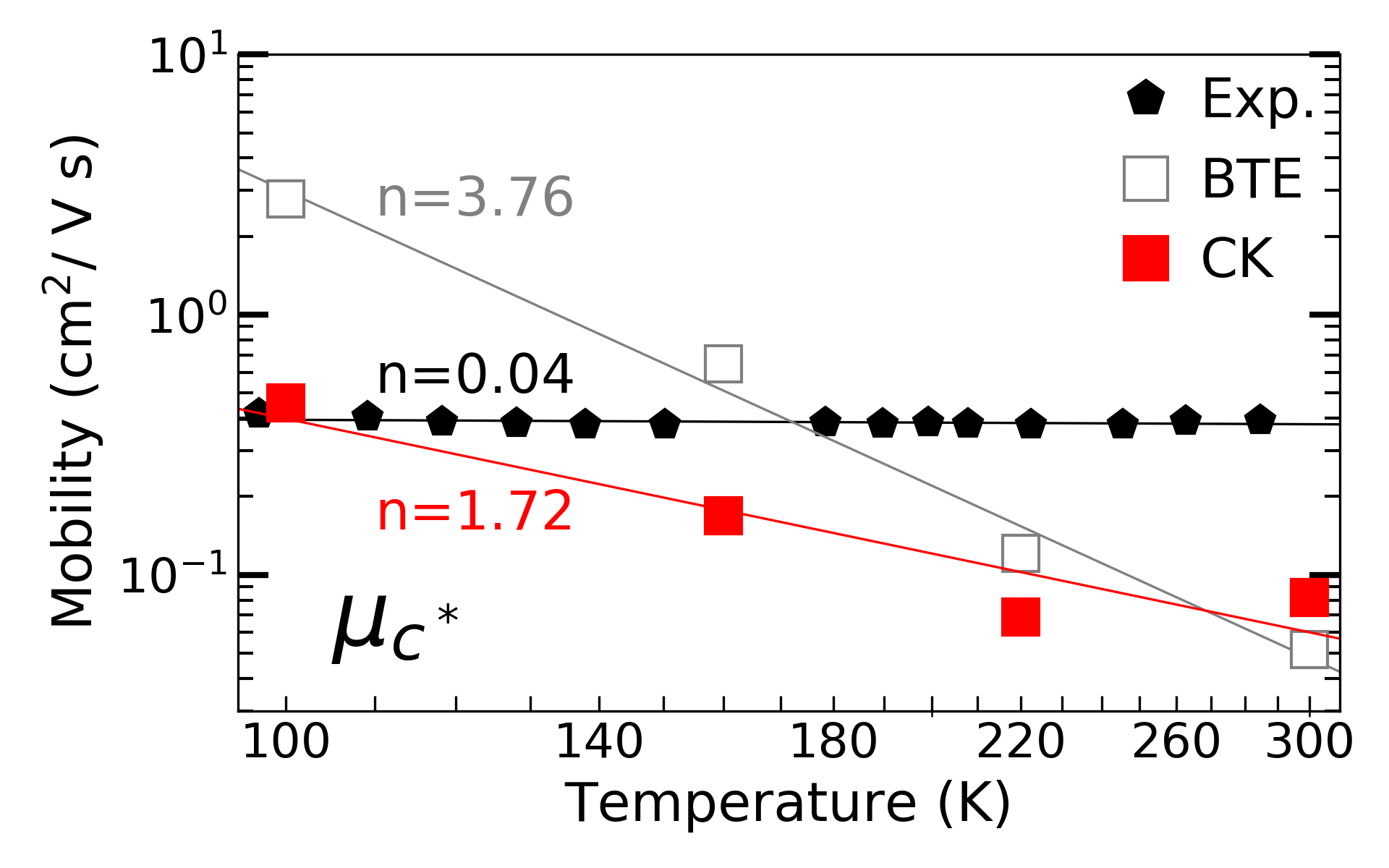} 
\caption{Electron mobility in the plane-normal $c^*$-direction in naphthalene. The plot compares CK and BTE calculations with experimental data from Ref.~\cite{madelung_naphthalene_nodate}.}
\label{fig:outofplane} 
\end{figure}

\vspace{-10pt}
In naphthalene, measurements of the mobility in the direction normal to the molecular planes ($c^*$ direction in Fig.~\ref{fig:structure}) point to a transport regime different from the in-plane directions. In experiments, the mobility along $c^*$ is lower than 1~cm$^2/\mathrm{Vs}$ and is nearly temperature independent between 
100$-$300~K~\cite{madelung_naphthalene_nodate,schein_band-hopping_1979,schein_observation_1978}. These trends suggest that charge transport normal to the molecular planes may occur in the small-polaron hopping regime, where the carriers are strongly localized and the e-ph interaction is so strong that a diagram-resummation technique such as the cumulant method is not expected to give accurate results.
\\
\indent
We calculate the plane-normal mobility using both the BTE and CK methods, and compare the results with experiments in Fig.~\ref{fig:outofplane}. 
The computed mobility decreases with temperature in both the CK and BTE approaches, deviating substantially from the nearly temperature independent mobility found in experiment. 
It is encouraging that the CK mobility agrees well with experiment at 100~K and its temperature dependence is weaker than in the BTE $-$ 
fitting the temperature dependence with a $T^{-n}$ power law gives an exponent $n=1.72$ in the CK and $n=3.78$ in the BTE method, versus $n=0.04$ in experiment. However, although the CK provides a significant improvement over the BTE, it is clear that neither method can accurately describe charge transport normal to the molecular planes.
\\
\indent
The electron bands in the GW band structure are very narrow in the plane-normal c*-direction [$\Gamma$ $\!-\!$ Z direction in Fig.~\ref{fig:cbmspectral}(b)], with large effective masses of order 15\,$m_e$ for the GW calculation done on the 100~K structure, and greater at higher temperatures. Combined with the absence of a power-law temperature trend in the experimental mobility, this relatively flat band suggests that electrons are nearly localized to a single molecular plane and that transport in the plane-normal direction occurs via small-polaron hopping. The failure of the CK approach in this regime highlights the need for predictive first-principles approaches to study charge transport in the small-polaron hopping regime in OMCs.\\
\newpage
\section*{\label{sec:conclusion} Conclusion}
\vspace{-10pt}
We studied the electron mobility in naphthalene crystal as a paradigmatic case of intermediate charge transport in OMCs. 
Combining a finite-temperature cumulant method and Green-Kubo transport calculations, we demonstrated accurate predictions of the electron mobility and its temperature dependence in the intermediate regime. 
%
%
Our results reveal a subtle interplay between inter- and intramolecular phonons: the low-energy intermolecular phonons limit the \mbox{mobility,} while the intramolecular modes are responsible for the formation of polarons and the resulting satellite peak in the spectral functions. 
The broad satellite removes spectral weight from the QP peak, modifying the mobility and its temperature dependence. By capturing these subtle polaron effects, our CK approach addresses the shortcomings of the BTE for modeling the intermediate transport regime. 
%
We also highlighted the limitations of the CK approach to describe polaron-hopping between molecular planes. Taken together, our work advances microscopic understanding of the intermediate transport regime and paves the way for accurate first-principles calculations of the carrier mobility in OMCs.\\

\section*{Methods}
\label{append:comp}
\vspace{-10pt}
\subsection*{First-principles calculations}
\vspace{-10pt}
We carry out first-principles density functional theory (DFT) calculations of the ground state and electronic structure of naphthalene using the \textsc{Quantum Espresso} code~\cite{giannozzi_quantum_2009,giannozzi_advanced_2017}. 
Thermal expansion of the lattice is taken into account by employing lattice constants~\cite{brock_temperature_nodate} taken from experiments at four different temperatures of 100, 160, 220, and 300~K. 
All calculations are carried out separately at these four temperatures. The initial atomic positions are also taken from experiment~\cite{grazulis_crystallography_2009,grazulis_crystallography_2012}. 
We use a kinetic energy cutoff of 90~Ry together with $2\times4\times2$ and $4\times4\times4$ $\mathbf{k}$-point grids for self-consistent and non-self-consistent calculations, respectively. The Grimme van der Waals correction~\cite{grimme_semiempirical_2006,barone_role_2009} is included during structural relaxation of the atomic positions. 
To improve the description of dynamical electronic correlations, we correct the DFT electronic band structure with $G_0W_0$ calculations, which include 500 bands in the polarization function and a cutoff of 10~Ry in the dielectric screening using the \textsc{Yambo} code~\cite{marini_yambo_2009}. The \textsc{Wannier90} code~\cite{mostofi_wannier90_2008} is employed to obtain Wannier functions and the corresponding transformation matrices, using the selected-columns-of-the-density-matrix method~\cite{pizzi_wannier90_2020}. The lattice dynamics and e-ph perturbation potential are computed with density functional perturbation theory (DFPT)~\cite{baroni_phonons_2001} calculations on a $2\times4\times2$ $\mathbf{q}$-point grid (here and below, $\mathbf{q}$ is the phonon wavevector). Using our \textsc{Perturbo} code~\cite{zhou_perturbo_2020}, the electron and phonon data are combined to form the e-ph coupling matrix elements~\cite{zhou_perturbo_2020}:
\begin{eqnarray}
\label{eq:g}
g_{mn\nu}(\mathbf{k},\mathbf{q})=\sqrt{\frac{\hbar}{2\omega_{\nu\mathbf{q}}}}\sum_{\kappa\alpha}\frac{\mathbf{e}^{\kappa\alpha}_{\nu\mathbf{q}}}{\sqrt{M_\kappa}}\langle m\mathbf{k+q}|\partial_{\mathbf{q}\kappa\alpha}V|n\mathbf{k}\rangle,\nonumber\\
\end{eqnarray}
where $|n\mathbf{k}\rangle$ are electronic Bloch states, $\omega_{\nu\mathbf{q}}$ are phonon energies, $\partial_{\mathbf{q}\kappa\alpha}V$ are e-ph perturbation potentials, $\mathbf{e}^{\kappa\alpha}_{\nu\mathbf{q}}$ are phonon displacement vectors, and $M_\kappa$ is the mass of atom $\kappa$. The absolute value of the gauge-invariant e-ph coupling strength shown in Fig.~\ref{fig:cbmspectral}(b) is computed for each phonon mode $\nu$ and phonon wavevector $\mathbf{q}$ as~\cite{zhou_perturbo_2020}
\begin{equation}
|g_\nu(\mathbf{k}\!=\!0,\mathbf{q})|=\sqrt{\sum_{mn}|g_{mn\nu}(\mathbf{k}\!=\!0,\mathbf{q})|^2/N_b}\,,
\end{equation}
where $N_b$ is the number of selected bands. The mobility calculations use a fine $\mathbf{k}$-grid of $60\times60\times60$ for the BTE and $30\times30\times30$ for the CK method. Both methods use between 10$^5$$-$10$^6$ randomly selected $\mathbf{q}$-points.\\
\subsection*{Electron-phonon scattering rate}
\vspace{-10pt}
The relaxation time $\tau_{n\mathbf{k}}$ used in the BTE is computed as the inverse of the scattering rate, defined as~\cite{bernardi_first-principles_2016}
\begin{eqnarray}
\label{eq:Gamma}
\Gamma_{n\mathbf{k}}(T)=&&\frac{2\pi}{\hbar}\sum_{m\nu\mathbf{q}}|g_{nm\nu}(\mathbf{k},\mathbf{q})|^2\nonumber\\
&&\times[(N_{\nu\mathbf{q}}+1-f_{m\mathbf{k+q}})\delta(\varepsilon_{n\mathbf{k}}-\varepsilon_{n\mathbf{k+q}}-\omega_{\nu\mathbf{q}})\nonumber\\
&&+(N_{\nu\mathbf{q}}+f_{m\mathbf{k+q}})\delta(\varepsilon_{n\mathbf{k}}-\varepsilon_{n\mathbf{k+q}}+\omega_{\nu\mathbf{q}})],
\end{eqnarray}
where $f_{n\mathbf{k}}$ and $N_{\nu\mathbf{q}}$ are electron Fermi-Dirac and phonon Bose-Einstein occupations in thermal equilibrium, respectively.\\ 

\vspace{10pt}
\subsection*{Cumulant method}
\vspace{-10pt}
The cumulant ansatz assumes that the retarded electron Green's function in the time domain takes the exponential form in Eq.~(\ref{eq:green}), where the cumulant function $C_{n\mathbf{k}}$ is defined as~\cite{kas_cumulant_2014,zhou_predicting_2019}
\begin{eqnarray}
C_{n\mathbf{k}}(t,T)=\int^{\infty}_{-\infty}d E \frac{|\mathrm{Im}\Sigma_{n\mathbf{k}}(E + \varepsilon_{n\mathbf{k}},T)|}{\pi E^2}(e^{-i E t} + i E t - 1).\nonumber\\
\end{eqnarray}
Here, $\varepsilon_{n\mathbf{k}}$ is the electron band energy, $E$ the electron energy, and $\Sigma_{n\mathbf{k}}$ is the lowest-order (Fan-Migdal) e-ph self-energy~\cite{mahan}:
\begin{eqnarray}
\label{eq:selfenergy}
&&\Sigma_{n\mathbf{k}}(E,T)=\sum_{m\nu\mathbf{q}}|g_{mn\nu}(\mathbf{k},\mathbf{q})|^2\nonumber\\
&&\times \left[\frac{N_{\nu\mathbf{q}}+f_{m\mathbf{k+q}}}{E-\varepsilon_{m\mathbf{k+q}}+\omega_{\nu\mathbf{q}}+i\eta}+\frac{N_{\nu\mathbf{q}}+1+f_{m\mathbf{k+q}}}{E - \varepsilon_{m\mathbf{k+q}}-\omega_{\nu\mathbf{q}}+i\eta}\right]\nonumber\\
\end{eqnarray}
whose temperature dependence is due to the occupation factors $N_{\nu\mathbf{q}}$ and $f_{n\mathbf{k}}$. Our cumulant Green's function includes e-ph Feynman diagrams of all orders: it sums over all the improper diagrams in which, at order $n$, the Fan-Migdal e-ph self-energy is repeated $n$ times and weighted by a $1/n!$ factor~\cite{mahan}. After Fourier-transforming the retarded Green's function in Eq.~(\ref{eq:green}) to the energy domain, we obtain the electron spectral function using Eq.~(\ref{eq:spectral}). 
%
We compute $\mathrm{Im}\Sigma_{n\mathbf{k}}(E)$ off-shell, using a fine energy $E$ grid, and $\mathrm{Re}\Sigma_{n\mathbf{k}}$ on-shell at the band energy $\varepsilon_{n\mathbf{k}}$, and use them as input to obtain the spectral function $A_{n\mathbf{k}}$~\cite{zhou_predicting_2019} as a function of electron energy $E$. 
Due to the exponential form of $G^{\mathrm{R}}_{n\mathbf{k}}$, the cumulant Green's function includes contributions from higher-order e-ph Feynman diagrams~\cite{mahan}.

\subsection*{Dyson-Migdal spectral function}
\vspace{-10pt}
The Dyson-Migdal (DM) spectral function is given by
\begin{eqnarray}
\label{eq:dyson}
A_{n\mathbf{k}}^{\mathrm{DM}}(E,T)=\frac{-\mathrm{Im}\Sigma_{n\mathbf{k}}(T)}{[E-\varepsilon_{n\mathbf{k}}-\mathrm{Re}\Sigma_{n\mathbf{k}}(T)]^2+[\mathrm{Im}\Sigma_{n\mathbf{k}}(T)]^2},\nonumber\\
\end{eqnarray}
\noindent
where $\Sigma_{n\mathbf{k}}(T)$ is the lowest-order e-ph self-energy [Eq.~(\ref{eq:selfenergy})] computed on-shell at the band energy $\varepsilon_{n\mathbf{k}}$. The DM spectral function has a Lorentzian shape as a function of energy, with a linewidth of $2\,\mathrm{Im}\Sigma_{n\mathbf{k}}(T)$ which is proportional to the e-ph scattering rate in Eq.~(\ref{eq:Gamma}),  $\Gamma_{n\mathbf{k}}(T)=2\,\mathrm{Im}\Sigma_{n\mathbf{k}}(T)/\hbar$~\cite{bernardi_first-principles_2016}.

\section*{Acknowledgements}
\vspace{-10pt}
This work was supported by the National Science Foundation under Grant No. DMR-1750613. J.-J.Z. acknowledges support from the Joint Center for Artificial Photosynthesis, a DOE Energy Innovation Hub, as follows: The development of some computational methods employed in this work was supported through the Office of Science of the US Department of Energy under Award No. DE-SC0004993. N.-E. Lee was supported by the Air Force Office of Scientific Research through the Young Investigator Program, Grant FA9550-18-1-0280. 
This research used resources of the National Energy Research Scientific Computing Center (NERSC), a U.S. Department of Energy Office of Science User Facility located at Lawrence Berkeley National Laboratory, operated under Contract No. DE-AC02-05CH11231.

\section*{Author Contributions}
\vspace{-10pt}
B.K.C. and M.B. conceived and designed the research. B.K.C. performed calculation and analysis. J.-J.Z. and N.-E.L. provided technical and theoretical support. M.B. supervised the entire research project. All authors discussed the results and contributed to the manuscript.

\section*{Competing Interests}
\vspace{-10pt}
The authors declare no competing interests.

\section*{Additional Information}
\vspace{-10pt}
\noindent
\textbf{Supplementary information} The online version contains supplementary information available at [insert URL].\\
\textbf{Correspondence} and requests for materials should be addressed to M.B.
\newpage


%

\end{document}